# What Limits Robustness in Deep Image Watermarking: An Analysis of Mechanisms and Their Scaling Across Capacities

**MARTA BISTROŃ[1], ZBIGNIEW PIOTROWSKI[1]**

[1]Institute of Communication Systems, Faculty of Electronics, Military University of Technology, 00-908 Warsaw, Poland

Corresponding author: Marta Bistroń (marta.bistron@wat.edu.pl)

This research was funded by the Military University of Technology, Faculty of Electronics, under grant no. EL_UGBWEL_12022026_04, titled "Neural methods of signal processing in spectrum monitoring systems, unmanned aerial vehicle detection, and data transmission protection."

**ABSTRACT** Robustness remains the principal open problem in deep image watermarking, and what limits it becomes sharper as payload grows. This paper asks whether capacity is itself the limit or only makes other limits visible, and answers in two parts. The first organizes the distortions a watermark must survive and the strategies developed to resist them, ordering each by the axis that governs it: payload capacity for the distortions, differentiability for the strategies. The second identifies and measures three mechanisms that limit robustness in schemes mapping the payload onto a spatial block grid with extraction trained separately from a frozen embedder: desynchronization of the payload grid, the resistance of codec-induced distortion to training, and the narrowing of the usable embedding-strength window. Payloads from 64 to 16384 bits are measured, well beyond the range those strategies address. Training the extraction stage against a codec proves not merely ineffective but harmful, degrading the reading at the operating points used in training. The limits follow the class of distortion rather than capacity itself, and none is removed by further training on the extraction side, because all three arise before extraction. An evaluation protocol making claims of generalization verifiable is also contributed. The conclusions are properties of a class of designs rather than of one implementation.

**INDEX TERMS** Bit error rate, deep learning, digital watermarking, payload capacity, robustness.

## I. INTRODUCTION

Recent advances in generative artificial intelligence have significantly transformed the creation, modification, and distribution of digital visual content [1], [2]. Modern diffusion models, generative adversarial networks, and large-scale multimodal systems enable the generation of highly realistic images that are often indistinguishable from authentic content. While these technologies create new opportunities in media production, education, and creative industries, they also introduce substantial challenges related to content authenticity, misinformation, copyright protection, and the proliferation of deepfakes. As a result, ensuring trustworthiness and traceability of digital media has become an increasingly important research topic.

Digital watermarking has emerged as one of the most promising technologies for protecting digital media in the era of artificial intelligence. By embedding imperceptible information directly into visual content, watermarking can support copyright protection, ownership verification, content authentication, and provenance tracking [3], [4]. These capabilities are increasingly recognized as essential components of trustworthy artificial intelligence frameworks, where mechanisms enabling content attribution and authenticity verification are required to ensure transparency and accountability in digital ecosystems.

This direction is reflected in standardization work. The specifications developed within the Coalition for Content Provenance and Authenticity define how signed metadata describing the origin and processing history of a media file are attached to that file [5]. Such solutions have an important limitation: metadata stored in the file container are lost on re-encoding, screen capture, format conversion, or publication through a service that strips unrecognized fields. A watermark embedded in the image content itself is considerably more durable under such operations, which is why the two approaches are regarded as complementary: metadata declare the provenance of the content, while the watermark carries that declaration through channels in which the file container does not survive.

The role a watermark can play in such a provenance chain depends directly on its information capacity. With a payload of a few dozen bits, the watermark is in practice an identifier pointing to an external database, so its usefulness ends the moment access to that database is lost. As capacity grows it becomes possible to move a complete description into the image itself: author, license, processing history, and a cryptographic signature. The file then becomes self-contained.

Reliable recovery under content transformations, however, remains the principal open problem. Digital media are routinely subjected to processing operations, compression, editing, and AI-driven modification, all of which may affect the embedded signal [6]. Raising capacity sharpens the difficulty, because it reduces the image area available to carry each payload element and with it the margin on which extraction relies. What has not been established is how the individual mechanisms behind robustness behave as that margin narrows. Whether the narrowing is itself what limits robustness, or whether the limits lie elsewhere and capacity only makes them visible, is the question this paper sets out to answer.

The paper is accordingly organized in two parts of comparable weight. The first systematizes the distortions a watermark must survive and the strategies developed to resist them, ordering each by the axis that governs it: capacity for the distortions, differentiability for the strategies. The second measures how the resulting limitations behave as capacity is raised well beyond the payloads for which those strategies were designed. The first part supplies the two axes along which the second is read.

The contributions of this paper are the following.

- A systematization of the field: distortions organized by threat model and by what each class damages, then mapped onto the payload capacity at which each becomes decisive. Robustness mechanisms are grouped into four categories rather than the usual three, with decoupled training isolated as a mechanism in its own right, and ordered by whether the model of the distortion must be differentiable.
- The identification and measurement of three mechanisms that limit robustness: synchronization of the payload grid under geometric transformation, trainability of codec-induced distortion while the embedder is frozen, and the width of the usable embedding-strength window. Each is stated as a property of a class of designs, then measured across payload capacities spanning more than two orders of magnitude.
- An evaluation procedure that makes claims of generalization verifiable. Every attack carries an explicit parameter, its measured severity is reported alongside the bit error rate, and its training status is annotated on a three-valued scale separating genuinely held-out distortions from those tested within a trained class at different parameters. Most operating points are those reported by widely used systems and are anchored to the work reporting them, so results can be set beside published figures.

The three mechanisms follow from two properties of the design, a payload mapped onto a spatial block grid and an extraction stage trained separately from a frozen embedder, rather than from any particular network architecture. The conclusions therefore transfer to other schemes sharing these two properties, and the system used in the experiments serves as a measurement apparatus rather than as the subject of the study.

The remainder of the paper is organized as follows. Section II discusses the categories of distortions affecting watermarking systems and relates them to payload capacity. Section III groups four categories of robustness enhancement strategies and orders them by their differentiability requirement. Section IV describes the evaluation protocol, Section V reports the three limiting mechanisms, Section VI discusses their design implications, and Section VII concludes.

## II. ROBUSTNESS CHALLENGES IN DEEP WATERMARKING

Distortions affecting digital watermarking systems may originate from both unintentional processing operations and deliberate attacks. This distinction is of practical importance, because it leads to two different threat models. Model-independent degradations, such as compression, format conversion, rescaling to the requirements of a service, or re-encoding in a transmission chain, are an unavoidable consequence of normal content exploitation and occur on a mass scale, regardless of whether anyone is interested in removing the watermark. Targeted attacks, by contrast, are designed with full awareness that a watermark is present, and often with knowledge of the algorithm used. Robustness against the former determines the practical usefulness of a system; robustness against the latter determines its security. From a technical perspective, these distortions can be categorized into signal processing operations, geometric transformations, video compression, and modifications performed using artificial intelligence.

### *A. SIGNAL PROCESSING ATTACKS*

Signal processing attacks constitute one of the most common categories of distortions considered in digital watermarking research. Unlike geometric transformations, these operations primarily modify pixel intensities while preserving the overall spatial structure of the image. Typical examples include lossy compression, filtering, and noise contamination.

Among them, JPEG compression remains the most frequently used benchmark due to its widespread adoption in image storage, transmission, and online content sharing.

Additional distortions such as Gaussian blur, median filtering, and additive noise are commonly applied to simulate quality degradation introduced by acquisition devices, communication channels, or post-processing operations. Although these transformations often have limited impact on visual perception, they may significantly alter the embedded watermark signal and reduce extraction accuracy. Consequently, robustness against signal processing distortions is widely regarded as a fundamental requirement for practical watermarking systems [7], [8].

What this class damages is the amplitude of the embedded signal relative to the cover, and it does so selectively in frequency. Lossy compression and low-pass filtering discard low-amplitude high-frequency content, which is precisely where an imperceptible watermark is most cheaply placed; this is why JPEG is commonly modeled during training as a suppression of high-frequency transform coefficients [9]. The allocation that minimizes visibility is at the same time the allocation most exposed to this class of attack, so robustness can be gained only by increasing embedding strength. This makes signal processing the class in which the trade-off between transparency and robustness is least avoidable. A practical consequence is that reported severities are poorly comparable between papers: nominally identical settings such as "JPEG quality 50" correspond to different amounts of degradation depending on the implementation and on the image content, so a bit error rate quoted without the measured severity of the attack conveys limited information. The same concern motivated the introduction of standardized robustness benchmarks for watermarking, which report detection performance jointly with the induced quality degradation [10].

### B. GEOMETRIC ATTACKS

Geometric attacks modify the spatial structure of an image rather than its pixel values. Typical examples include cropping, resizing, rotation, scaling, and translation. Unlike signal processing distortions, these transformations do not necessarily weaken the watermark signal; they displace the spatial reference against which extraction reads it. Reliable recovery therefore becomes a problem of synchronization rather than of signal strength, and this is what makes the class qualitatively different from the preceding one.

Among geometric attacks, cropping and resizing are particularly important because they frequently occur during image editing, content reposting, and adaptation to different display formats. Rotation and scaling introduce additional difficulty by changing the spatial arrangement of the features used for extraction. Geometric robustness consequently remains one of the least satisfactorily resolved requirements in modern watermarking [11].

The strongest resistance to geometric transformations is reported by schemes that carry no message at all. Where the watermark is embedded as a structured pattern in a transform domain of a diffusion model's initial noise, geometric invariance follows from the construction of the pattern rather than from training [12]. That same construction, however, bounds the payload: later work in this line has been observed either to remain vulnerable to geometric attacks or to lack the capacity to embed sufficient multi-bit information [13]. A representative example is ZoDiac, presented as an attack-resilient diffusion-based watermark, which states explicitly that it focuses on zero-bit watermarking (hiding and detecting a mark) and defers the encoding of an actual message to future work [14]. Its cost is also considerable, as watermarking a single image requires per-image latent optimization taking from 45 to 256 seconds. Even at that cost, the authors report that all evaluated methods but one fail under rotation, which they attribute to the non-rotation-invariant nature of the inversion process. Multi-bit schemes with meaningful geometric robustness do exist, but they operate at payloads of the order of tens to about one thousand bits and obtain that robustness either from a representation in which invariance is built in, or from explicit augmentation during training.

### C. VIDEO COMPRESSION AS A REAL DISTORTION CHANNEL

Video compression performed by codecs such as H.264/AVC [15] and H.265/HEVC [16] deserves separate discussion. It is arguably the most frequently encountered degradation channel for content in circulation, and at the same time the channel most difficult to reproduce during training. Hybrid codecs combine intra- and inter-frame prediction, block transforms, quantization controlled by a quality parameter, entropy coding, and in-loop reconstruction filters, including deblocking and, in HEVC, sample adaptive offset. The resulting distortion cannot be reduced to a composition of a few simple differentiable operations, and its characteristics depend on image content, the operating point of the encoder, and the group-of-pictures structure.

This property has direct consequences for how robustness can be built. In the classical training scheme with simulated attacks, the distortion must be differentiable because it lies on the gradient propagation path; a real codec does not satisfy this condition. This issue, which forms the cross-cutting axis of Section III, is particularly relevant for architectures derived from video sequence processing, for which compression is a natural rather than hypothetical deployment scenario [17], [18].

It is worth noting that JPEG compression and hybrid video codecs belong to the same family of methods. Both quantize the coefficients of a transform computed on image blocks and introduce artifacts of the same kind, namely a visible block structure and the loss of high-frequency components. They differ in block size, the form of the transform, and the presence of prediction, but the underlying mechanism of signal degradation is shared. Whether robustness acquired against one compression of this family carries over to another is therefore a question about the structure of the signal rather than an arbitrary hypothesis.

TABLE 1
DISTORTION CLASSES AND THEIR DEPENDENCE ON PAYLOAD BLOCK SIZE IN SCHEMES THAT MAP THE PAYLOAD ONTO A SPATIAL BLOCK GRID.

| Distortion class | Mode of degradation | Dependence on block size |
|---|---|---|
| Signal processing | frequency-selective attenuation of amplitude | a smaller block is a higher-frequency feature, so the payload moves into the band this class removes |
| Geometric | displacement of the spatial reference | displacement tolerance is bounded by block size, so it falls as blocks are subdivided |
| Video compression | content-adaptive attenuation of amplitude | a payload block smaller than the transform block can no longer be quantized independently |
| AI-based | replacement of the carrier | the payload is resynthesized rather than degraded, so block size is not the governing quantity |

What this class damages is again the high-frequency, low-amplitude part of the signal, but unlike a fixed filter it does so adaptively. Rate control allocates bits according to content, so the same nominal quality parameter produces different degradation on different images, and the variance across images is correspondingly large, a property that averaged error rates conceal. Two further characteristics distinguish this class. First, it is the class in which the differentiability requirement discussed in Section III binds most tightly, since no compact differentiable expression reproduces the behavior of a hybrid codec. Second, although video is one of the dominant distribution channels for images in practice, evaluation against real codecs remains uncommon in the image-watermarking literature; among the reference systems, it is chiefly the video-oriented ones that report it [19].

### D. ATTACKS BASED ON ARTIFICIAL INTELLIGENCE

The rapid development of generative artificial intelligence has introduced a new class of threats to digital watermarking systems. Unlike conventional distortions, AI-based attacks can modify or regenerate image content while preserving high perceptual quality, making watermark detection and extraction significantly more challenging.

Examples of such attacks include diffusion-based image editing, image inpainting, content regeneration, and style transfer techniques. Recent studies have also demonstrated the possibility of using neural networks specifically trained to detect and remove embedded watermarks, as well as adversarial perturbations designed to disrupt watermark extraction processes. Furthermore, latent-space manipulations performed within diffusion and variational autoencoder models may effectively eliminate watermark information without introducing visually noticeable artifacts [20], [21].

What this class damages is neither the amplitude of the signal nor its synchronization, but the carrier itself. A regenerated image is not a degraded version of the watermarked one; it is a different image that preserves the semantics of the original. Robustness in the signal-processing sense is therefore not the appropriate notion here, and the relevant question becomes whether a mark can survive a semantics-preserving re-synthesis of the content. This class also differs from the others in how it evolves over time. The severity of JPEG compression or of a Gaussian filter is bounded by the operation itself, whereas the strength of a regeneration attack scales with the quality of the underlying generative prior, and therefore increases without any change in the attacker's intent or effort.

The emergence of these attacks highlights the limitations of robustness mechanisms focused solely on conventional distortions and motivates the development of more generalized approaches capable of preserving watermark information under previously unseen transformations.

### E. PAYLOAD CAPACITY AS A SCALING PARAMETER

The four classes above were described without reference to how the payload is arranged inside the image. For many schemes that is sufficient, because the payload is spread across the whole image and read globally. This paper is concerned with schemes that map the payload onto a spatial block grid, where each block carries one payload element. In such schemes the capacity fixes the block size, and the block size governs how each class of distortion affects the reading. Table 1 states the relation class by class.

Capacity therefore does not raise difficulty uniformly. It governs three of the four classes, while in the fourth it does not enter in the same form. The same increase in payload thus sharpens some limitations and leaves others unchanged. This is the axis along which the present study is organized.

The classical statement of the field is a three-way trade-off between payload, imperceptibility and robustness [22], [23]. What enters it is the signal carried by each payload element, amplitude multiplied by area, rather than the number of bits. A block grid separates the two, since the payload can be raised either by subdividing blocks or by resolving more levels within one, and Section V uses that separation throughout.

## III. EVOLUTION OF ROBUSTNESS ENHANCEMENT METHODS

The distortions described above have given rise to four robustness strategies. The earliest simulated the attack during training, resisting distortions defined in advance. Later work sought a more faithful reproduction of complex degradations, compression in particular. A third line separates the training of embedding from the training of robustness into distinct stages. The most recent shifts the emphasis from modeling the distortion to learning an invariant representation. The four are discussed in turn and then compared against a common set of criteria.

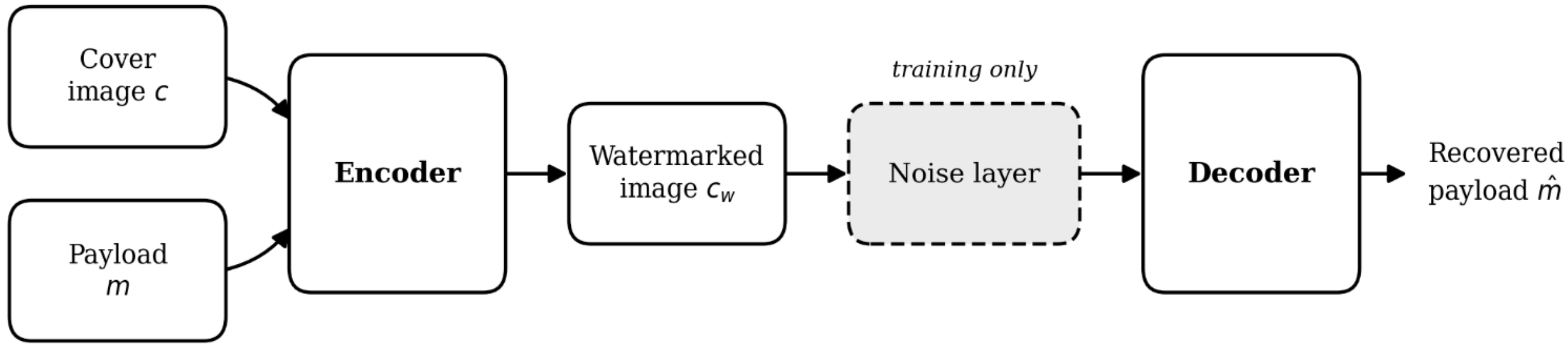

**FIGURE 1.** Encoder–noise layer–decoder pipeline used in noise-layer-based training. The noise layer is present only during training.

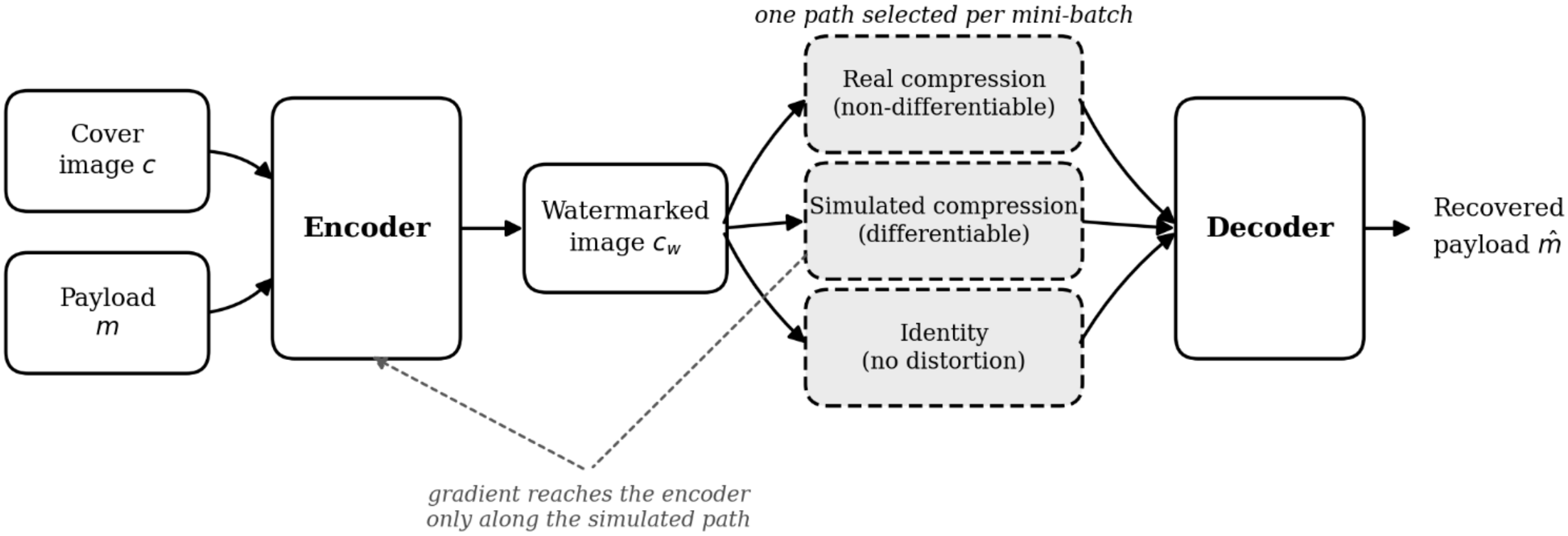

**FIGURE 2.** Path selection per mini-batch in learned distortion models. One of three paths, real compression, a differentiable surrogate, or identity, is selected at random for each mini-batch during training. Gradient reaches the encoder only along the simulated path.

### A. NOISE-LAYER-BASED TRAINING

Noise-layer-based training is the earliest and still the most widely adopted strategy for improving robustness in deep learning-based watermarking. Since the emergence of the first end-to-end architectures, robustness has commonly been achieved by incorporating an attack-simulation module directly into the training process. The main idea is to introduce differentiable approximations of common distortions, enabling the decoder to learn resistance against expected degradations.

This paradigm was popularized by early deep watermarking frameworks, most notably HiDDeN [6], and remains a fundamental component of contemporary systems such as TrustMark [24], deployed for provenance at arbitrary image resolution. A typical architecture, shown in Fig. 1 consists of an encoder responsible for watermark embedding, a noise layer simulating distortions, and a decoder recovering the embedded information. During training, watermarked images are subjected to operations such as JPEG compression, blur, cropping, resizing, or additive noise before reaching the decoder. By exposing the network to these transformations, the model gradually learns features that are less sensitive to the corresponding distortions [6].

The popularity of this approach stems from its simplicity, compatibility with end-to-end optimization, and effectiveness against scenarios anticipated at design time. It has, however, two limitations. The first, widely acknowledged, is that the resulting robustness depends on the set of distortions used during training, so transformations outside that set may substantially reduce extraction accuracy [25]. The second follows from where the noise layer sits. Because it lies on the gradient propagation path between the encoder and the decoder, every operation it contains must be differentiable, and this requirement, rather than any property of the network, decides which degradation channels may enter training at all. As shown in Section II, the compressions encountered most often in practice do not meet it.

### B. LEARNED DISTORTION MODELS AND CODEC SIMULATION

Although noise-layer-based training effectively improves robustness against predefined distortions, faithfully reproducing complex real-world transformations remains difficult. Two responses to this difficulty have been proposed, both of which keep the training process in a single stage.

Phase 1: joint training

Encoder → stego image → Decoder

encoder and decoder trained jointly, end to end

Phase 2: decoupled fine-tuning

Encoder (frozen) → Decoder (fine-tuned)

**FIGURE 3. Two-phase training procedure used in Video Seal and StyleMark. Phase 1 optimizes encoder and decoder jointly, as in noise-layer-based training (Fig. 1). Phase 2 freezes the encoder and fine-tunes the decoder alone, the decoupled scheme described in this section.**

Early attempts incorporated an actual compression channel directly into the training pipeline, allowing encoder and decoder to learn robustness against realistic codec-induced degradations [18]. Such approaches are effective, but they increase training complexity and remain dependent on the availability of the target processing chain. Subsequent work extended this concept by introducing trainable neural modules that approximate the behavior of the compression process itself [17]. Instead of executing the codec during optimization, a dedicated network learns the distortion characteristics from data and acts as a differentiable surrogate of the original transformation. Building such a surrogate requires no differentiation through the real distortion either. The network is fit offline, in a separate supervised step, to reproduce the output of the real operation on paired examples, so the real degradation is needed only to generate training targets in a forward pass, never as part of a gradient path.

A distinct variant of this strategy is the solution adopted in MBRS [9], illustrated in Fig. 2, where several paths are used alternately: for successive mini-batches, real compression, simulated compression, or a noise-free path is selected at random. The real compression provides fidelity of reproduction, while the simulated one, already mentioned in Section II-A, provides the gradient required to update the encoder. All solutions in this category share a common feature: they preserve the single-stage nature of training at the cost of fidelity of the distortion, because a surrogate remains an approximation of the operation it replaces.

### *C. DECOUPLING THE TRAINING OF EMBEDDING AND EXTRACTION*

A different answer to the same problem is given by a family of solutions in which, rather than making the distortion differentiable, it is removed from the gradient path of the encoder. This is achieved by separating training into two stages. In the first, encoder and decoder are trained jointly without distortions; in the second, the encoder is frozen and only the decoder is fine-tuned, this time on distorted images.

This scheme was formulated in [26] as a two-stage separable deep learning framework, composed of noise-free end-to-end adversarial training followed by noise-aware decoder-only training. The motivation given by the authors is the requirement identified above: under single-stage training the attack must be simulated in a differentiable manner, which is not always applicable in practice, whereas the decoder obtained in the second stage accepts any type of distortion. The consequence is fundamental. Since the gradient passes through the decoder alone, real JPEG compression, a real video codec, or any other operation not expressible analytically can be introduced into training without constructing a differentiable substitute for it.

The approach has been criticized. It was argued in [9] that separating the stages yields a solution that is only locally optimal, because the encoder never observes the distortion and therefore cannot adapt the spatial distribution of the embedded information to it. The objection is well founded and identifies the cost of that freedom in the choice of distortions.

The scheme has nevertheless not been abandoned, and it has recently reappeared in reference systems, not as a replacement for joint training but as a second stage added on top of it. Both Video Seal [19] and StyleMark [27] follow this two-phase procedure, shown in Fig. 3: the encoder and decoder are first optimized jointly, exactly as in Section III-A, and only afterward is a second, decoupled phase used to fine-tune the extractor with the embedder frozen. Video Seal reports improved extraction accuracy from this second phase, among others for sequences subjected to H.264 compression, at unchanged image quality. StyleMark follows the same pattern, building on what the authors describe as solid end-to-end training with a subsequent stage of decoder fine-tuning under random noise, to obtain robustness against generative style transfer. Decoupling should therefore be regarded not as an implementation variant of the preceding categories, but as a stage that complements standard joint training rather than substituting for it, with its own, separate profile of costs and benefits.

### *D. INVARIANT-DOMAIN LEARNING*

The fourth approach shifts the emphasis from modeling the distortion to learning a representation. Instead of teaching the network how individual attacks modify the image, the objective is to learn a latent representation that remains stable under multiple transformations of the same content. Differently distorted versions of a watermarked image are encouraged to yield similar feature representations, while unrelated images are mapped to distant regions of the latent space [28].

This objective is typically realized through a triplet-based training signal, originally proposed for metric learning [29]: an anchor sample, a positive sample obtained by distorting the anchor, and a negative sample drawn from unrelated content are jointly embedded, and the loss pulls the anchor and the positive together while pushing the negative apart. Applied to watermarking, the anchor is a stego image, the positive is a distorted version of it, and the negative is typically a different, unrelated stego image.

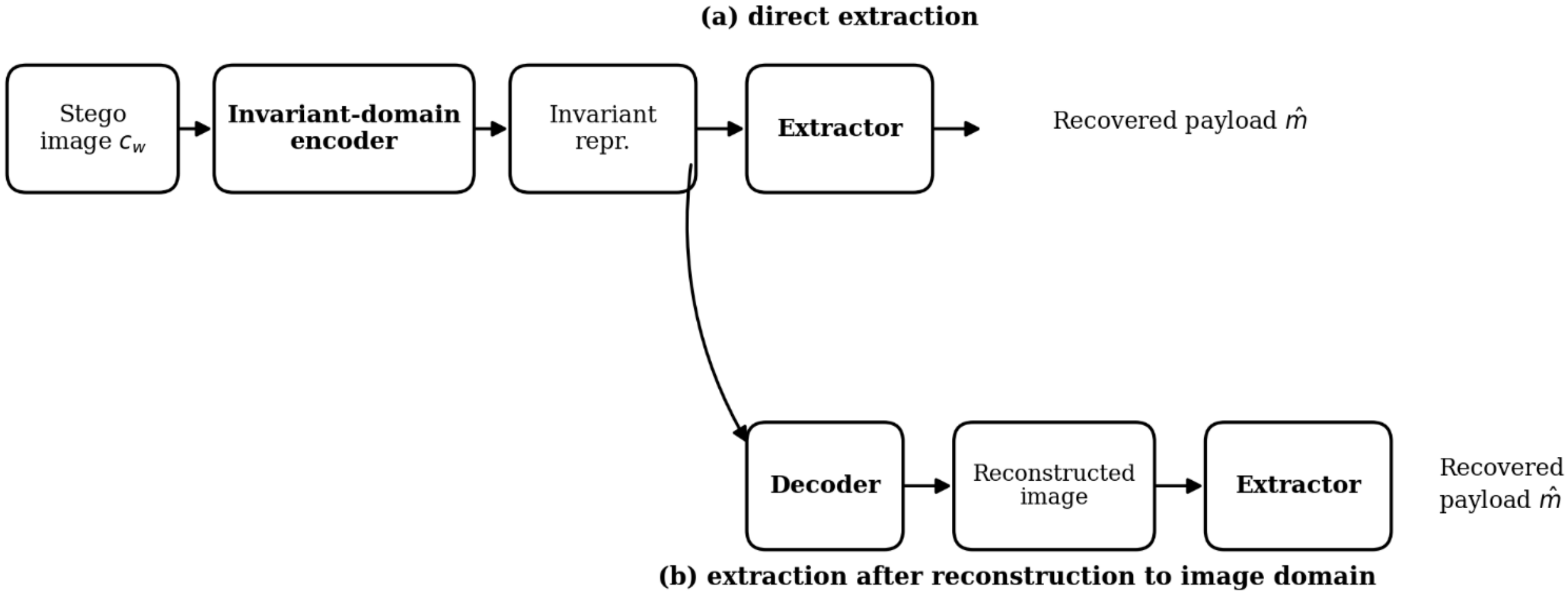

**FIGURE 4. Two strategies for extracting a watermark from an invariant-domain representation. (a) The message is decoded directly from the invariant representation, without reconstructing an image. (b) The representation is first decoded back into an image-domain reconstruction, and a conventional image-based extractor is then applied.**

Watermark extraction from such a representation can then proceed in one of two ways, shown in Fig. 4. In the first, the message is read directly from the invariant embedding, without ever reconstructing an image; this is the approach taken in self-supervised latent-space watermarking, where detection and decoding are performed within the same latent space in which the mark was embedded [30]. In the second, the invariant representation is first decoded back into an image-domain reconstruction, and a conventional, image-based extractor is then applied to that reconstruction; this is the path followed in [28], whose decoder reconstructs an image-domain tensor from the invariant representation before a separate extractor recovers the watermark. The two variants trade off differently: reading directly from the embedding avoids reconstructing an image the network was never explicitly trained to produce faithfully, while reconstructing first keeps the extractor compatible with architectures designed for pixel-domain input.

Two kinds of invariance should be distinguished at this point. In the schemes discussed in Section II-B, invariance follows from the construction of the embedded pattern and is therefore available without any training, at the cost of a bounded payload. Here, by contrast, invariance is learned, and the payload is not constrained by the geometry of a fixed pattern. The two are complementary rather than competing, and they fail in different ways.

In relation to the preceding category, one point is essential. Invariant-domain learning is decoupling taken one step further: the second stage is not merely fine-tuned, but equipped with a training objective of its own, formulated at the level of the representation. It therefore inherits the freedom in the choice of distortions described above and adds to it the promise of generalization beyond the set of transformations observed during training. It is at the same time the youngest of the four lines, with results of a preliminary character, which should be taken into account when assessing its maturity.

### E. DIFFERENTIABILITY AS A DESIGN CRITERION

The categories presented above differ in their robustness mechanism, yet they can be ordered along a single cross-cutting criterion: the differentiability requirement imposed on the model of the distortion. This criterion decides something more fundamental than effectiveness. It decides which degradation channels can be taken into account during training at all.

The clearest illustration is a comparison of two of the solutions discussed. In MBRS, a simulated variant of JPEG compression is constructed precisely because the distortion lies on the gradient path of the encoder and must remain differentiable. Under decoupled training such a substitute ceases to be necessary, because the distortion is an input to a separately trained stage rather than an intermediate layer. The comparison also shows that the two approaches incur their costs on opposite sides: single-stage training loses fidelity of the distortion, whereas decoupling loses co-adaptation between encoder and decoder. Neither cost is negligible, and neither is decided by the architecture of the network.

One boundary is shared by all four categories. As noted in Section II-D, attacks that regenerate image content do not degrade the signal but replace it, so robustness obtained by training on signal-level distortions does not transfer to them in any obvious way [20], [21]. This is what motivates evaluating robustness by generalization to distortions not observed during training, realized by separating the training and testing distortion sets, a convention adopted among others in [28] and [25].

### F. COMPARATIVE ANALYSIS

TABLE 2
COMPARISON OF ROBUSTNESS ENHANCEMENT CATEGORIES IN DEEP LEARNING-BASED WATERMARKING

| **Criterion** | **Noise layer** | **Learned distortion models** | **Decoupled training** | **Invariant domain** |
|---|---|---|---|---|
| Robustness mechanism | attack simulation | learned degradation modeling | extraction fine-tuning | representation learning |
| Training objective | resistance to simulated attacks (joint) | resistance to a learned degradation model (joint) | robust extraction under real distortion (frozen embedder) | a distortion-invariant representation (frozen embedder) |
| Differentiability of distortion required | yes | yes (surrogate) | no | no |
| Encoder observes the distortion | yes | yes | no | no |
| Dependence on the distortion set | high | moderate to high | moderate | low |
| Embedding and extraction co-adaptation | full | full | none | none |
| Research maturity | established | developing | established | early |
| Representative works | [6], [7], [8] | [9], [17], [18] | [19], [26], [27] | [28] |
| Payload in the representative works (bits) | 30–100 | 64–256 | 30–256 | 64 |

Taken together, the four categories trace a gradual transition from explicit modeling of the attack towards robustness built at the level of the representation, with an intermediate step, the separation of training stages itself, that is frequently omitted from comparative surveys. The key properties of the four categories are summarized in Table 2.

The rows concerning the required differentiability and the observation of the distortion by the encoder correspond to the criterion described in Section III-E, and make it possible to read off where the boundary between single-stage and decoupled solutions runs.

One property is common to all four categories and has not been stated so far. They were developed, and are reported, at payloads ranging from a few dozen to a few hundred bits, as Table 2 shows. The upper end has recently been pushed to 1024 bits by ChunkySeal, a scaled-up variant of one of the systems discussed above, which reports image quality and robustness essentially unchanged with respect to its 256-bit baseline [11]. That result is instructive for what it required: an embedder roughly ninety times larger and a move from luma-only to three-channel embedding, that is, more model and more carrier rather than a different robustness mechanism. Within these works robustness is reported as a single aggregate figure, and where capacity is varied, it is varied to show that a proposed method retains it, not to ask which of its mechanisms gives way first. How the individual mechanisms behave as capacity grows therefore remains open, and it is what the second part of this paper measures.

## IV. EVALUATION PROTOCOL

The measurements reported in Section V are obtained on one instance of the class of schemes defined in Section I: a payload mapped onto a spatial block grid, embedded by a convolutional encoder, and read by an extractor whose robustness, where it is built at all, is built in a stage separated from a frozen embedder. This section states only those properties of that instance which enter the arguments of Section V, together with the protocol under which it is measured. The mapping of payload bits onto the two-dimensional pattern is described in [31], and the training procedure and its stability in [32].

### A. PAYLOAD MAPPING AND OPERATING POINT

Payload bits are arranged into a two-dimensional pattern covering the cover image, so capacity fixes the size of the region carrying one payload element. On covers of 256 by 256 pixels the four capacities examined correspond to blocks of 32, 16, 8 and 2 pixels on a side. Each block carries one bit, so capacity and block area are the only quantities that change across the four variants: the architecture, the training data and the training procedure are identical. Variants are identified throughout by payload and bits per block, written as 64_1.

The capacity dependence reported in Sections V-A and V-C follows from this property alone, which is not particular to the scheme measured here. The adjustable subsquares algorithm divides the cover into square regions, gives each one symbol to carry, and treats the region size as a parameter the designer chooses: the same arrangement, on a different network and for video [17], [18]. Diffusion-based watermarks that write payload bits into spatially indexed carriers in a latent space index them the same way, and Section V-A shows that they fail the same way as well [13]. What follows is therefore a property of a construction, not of one implementation.

Embedding modifies luminance only. Distribution pipelines subsample chrominance by a factor of two in each dimension before any attack [15], [16], so a payload placed there is degraded by the container format itself. The reference systems of Section III make the same choice [11]. The opposite choice, made by schemes that carry the payload in chrominance [18], is defensible where invisibility weighs more than survival of the container format. Here the watermarked image is composed as the cover plus the luminance difference produced by the encoder, which leaves chrominance bit-exact and makes embedding strength a single

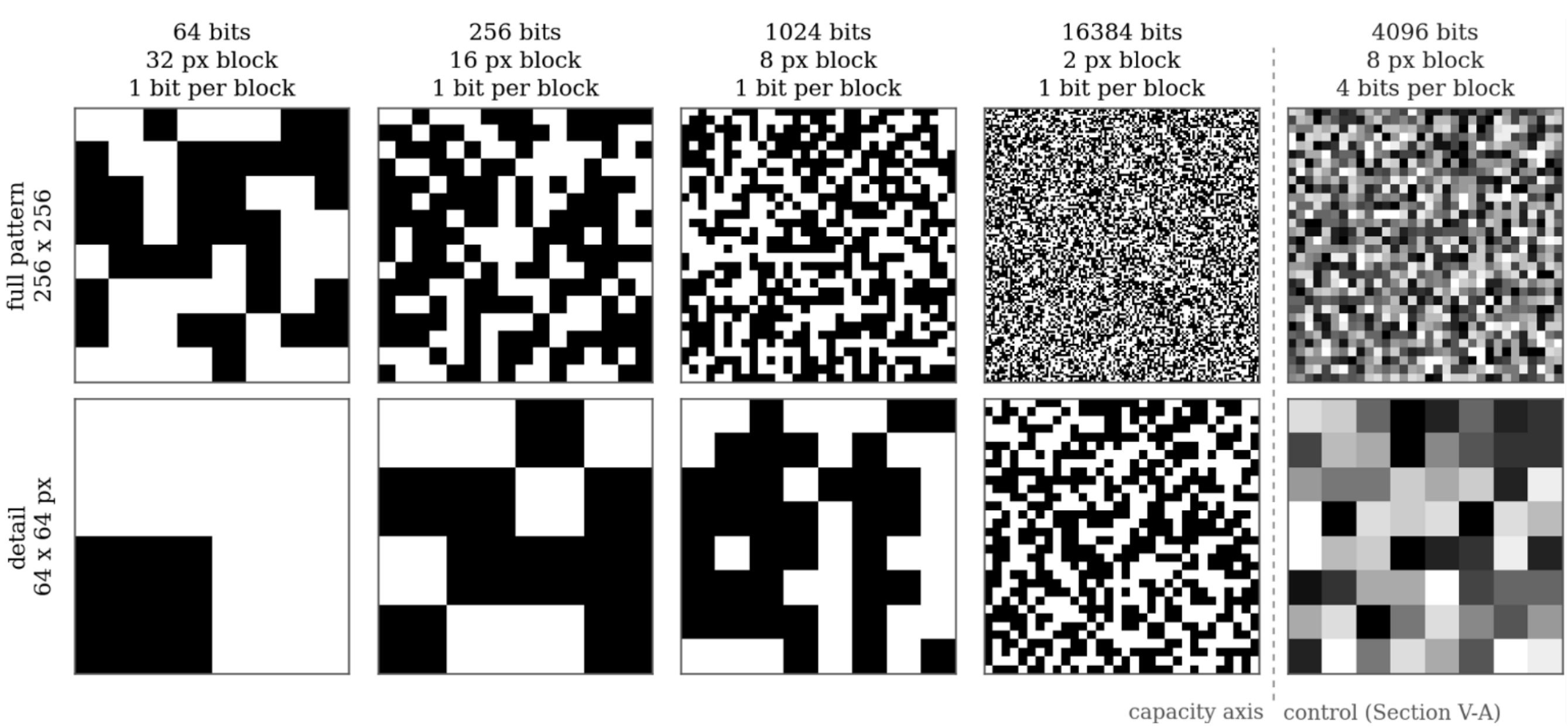

**FIGURE 5. Payload patterns for the capacities studied. Capacity fixes the block size, shown in the detail row. The rightmost variant carries four bits per block in its grey level and therefore shares its block size with the 1024-bit variant.**

TABLE 3
OPERATING POINTS OF THE FIVE VARIANTS STUDIED. COCO VAL2017, N = 1000.

| Variant (bits_bits per block) | Block size (px) | PSNR (dB) | SSIM | LPIPS |
|---|---|---|---|---|
| 64_1 | 32 x 32 | 39.07 | 0.9889 | 0.0055 |
| 256_1 | 16 x 16 | 39.51 | 0.9894 | 0.0027 |
| 1024_1 | 8 x 8 | 38.43 | 0.9855 | 0.0027 |
| 16384_1 | 2 x 2 | 36.12 | 0.9762 | 0.0042 |
| 4096_4 | 8 x 8 | 36.81 | 0.9790 | 0.0034 |

scalar, variable after training without retraining. That scalar is the subject of Section V-C.

The class defined in Section I also admits an extraction stage trained separately from a frozen embedder, in the sense of Section III-C. The variants measured in Sections V-A and V-C are taken before any such stage is applied. Encoder and extractor are trained jointly on undistorted images, and no distortion of any kind takes part in their training. Section V-B is the only place in this paper where the second, decoupled stage is introduced, and it is introduced there as the object of study rather than as part of the measurement setup.

A fifth variant is used once, in Section V-A, as a control. It encodes four bits per block in the grey level rather than in the sign, placing 4096 bits on the same grid of 8 pixel blocks as the 1024-bit variant. Fig. 5 shows the payload patterns of all five variants together: in the four that form the capacity axis the block shrinks as the payload grows, while the control repeats the block size of the 1024-bit variant and adds capacity in the grey level instead. Block size and capacity, which move together across the capacity axis, are therefore separated in that pair, which is what makes it useful there. The control lies outside the capacity axis under study.

Robustness is meaningless without the transparency at which it is obtained, so

Table 3 reports the operating point of each capacity, together with that of the control just described, measured on clean watermarked images as peak signal-to-noise ratio (PSNR), structural similarity (SSIM) and learned perceptual image patch similarity (LPIPS). All five were trained only to embed a payload and recover it from an undistorted image; no robustness objective and no simulated attack took part in producing them. Transparency falls by three decibels across a 256-fold increase in payload, and the small departure from monotonicity between the two lowest capacities reflects that the four variants were trained independently rather than any property of capacity itself. The figures are given for COCO val2017 [33]; the corresponding values on BSDS300 [34] are 1.0 to 1.6 dB higher at every capacity, so the two sets differ in level but agree on how the capacities compare.

Validation measurements are not reported anywhere in this paper, but they are not absent from the procedure: they select the best checkpoint and drive the early-stopping criterion during training, following [32]. Keeping them out of the results is what prevents that role from becoming an indirect leak, since a quantity used to choose a model cannot also serve as an unbiased estimate of it. The separation matters quantitatively as well: validation transparency runs 2 to 4 dB above the values in Table 3, because the training corpus is downsampled to the working resolution while both test sets are native crops at that resolution.

TABLE 4
THE ATTACK SUITE WITH PARAMETERS AND LITERATURE ANCHORS

| Class | Attack | Parameters | Anchor |
|---|---|---|---|
| Point operations | brightness | 0.5, 1.5 | TrustMark [24], Video Seal [19] |
| | contrast | 0.5, 2.0 | Dasgupta and Zhong [28] |
| | hue | 0.25 of the color circle | Dasgupta and Zhong |
| | saturation | 2.0 | Dasgupta and Zhong |
| | solarize | threshold 0.5 | Dasgupta and Zhong |
| | histogram equalize | luminance | Dasgupta and Zhong |
| Filtering and noise | Gaussian blur | sigma 1.0, 2.0; kernel 7 | HiDDeN [6], MBRS [9], TrustMark |
| | Gaussian noise | sigma 0.03, 0.06 | Dasgupta and Zhong, TrustMark |
| | salt-and-pepper | 1%, 5% | Dasgupta and Zhong |
| Compression | JPEG | quality 90, 75, 50, 30, 10 | MBRS, HiDDeN, Dasgupta and Zhong |
| | H.264 | CRF 23, 30 | Video Seal |
| | H.265 | CRF 28 | this work |
| Spatial | cutout | 20% of area blanked | Dasgupta and Zhong |
| | crop and rescale | 80%, 50% of area retained | Video Seal, ChunkySeal [11] |
| Geometric | rotation | 10, 30 degrees | Video Seal, ChunkySeal |
| | rescale | 0.7, 1.5 | TrustMark (extended range) |
| Composite | H.264, brightness, crop | CRF 30 · factor 0.5 · 50% of area retained | Video Seal (exact sequence) |
| | H.264, brightness, JPEG | CRF 30 · factor 0.5 · quality 75 | this work |
| Control | identity | | — |

### B. ATTACKS AND THEIR OPERATING POINTS

The suite defined for this protocol contains thirty-two attack points, grouped in Table 4 by what each does to the signal, following the classification of Section II. Every one carries an explicit parameter rather than a range from which a value is drawn, so a reported error rate belongs to a stated operating point and can be reproduced exactly. Wherever possible the parameter was chosen to coincide with a point reported in one of the works used as anchors in Sections II and III, so individual rows remain directly comparable with published figures rather than merely similar to them.

Three entries need more than a parameter to be read. The first is the blur, specified by kernel as well as by standard deviation: a 3 by 3 kernel truncates the Gaussian so severely that strength stops growing beyond roughly 1.0, so raising it from 1.0 to 2.5 moves the measured severity only from 30.13 to 28.98 dB, against 28.64 to 24.27 dB at a kernel of 9. A standard deviation alone therefore does not identify an operating point, and the kernel is stated throughout, here 7.

The second is compression, applied with real encoders rather than differentiable approximations: JPEG through an image library, H.264 and H.265 through a real codec. No operation in the protocol lies on a gradient path, so no substitute is required. This is the criterion of Section III-E applied to the measurement rather than to the design. The anchor gives the parameter, not the implementation: MBRS, for instance, uses both a real and a differentiable JPEG, and every attack here is the real one.

The third is the pair of composite attacks, which exists because several of the anchor works evaluate compositions rather than single operations. Video Seal reports them as a separate category, and for TrustMark a composition is the default evaluation. The first of the two reproduces the exact sequence used by Video Seal [19], so its figure is directly comparable with theirs. The second replaces the crop with JPEG at quality 75, so that it exercises everything the first does except displacement of the block grid. A single composite attack cannot separate the geometric mechanism of Section V-A from the rest. The pair can, which is why both are included.

### C. TRAINING STATUS AND MEASURED SEVERITY

Robustness to distortions not seen during training is the property most often claimed and least often verifiable. The usual practice is to divide distortions into a training set and a testing set and to report the second as held out, but the division is made by distortion class rather than by parameter, so a system trained on blur and tested on blur at a different width is still described as generalizing. The claim and the evidence then do not match.

Each attack therefore carries a three-valued annotation of its training status rather than a two-valued one: excluded from training entirely; applied within the range of parameters seen in training; or belonging to a class seen in training but applied outside that range. The third value is the one the two-valued convention conceals.

For the variants reported in Section V the annotation takes a single value. They are trained without any distortion at all, so every one of the thirty-two attack points is genuinely held out. Every number in Section V is a measurement of generalization to an unseen distortion, not a measurement of a condition it was trained for, and the mechanisms identified there are therefore not artifacts of a training set. The annotation becomes discriminating once a system trained with a distortion profile is evaluated on the same suite, which is what Section V-B does.

TABLE 5
ATTACKS MATCHED ON MEASURED SEVERITY AND DIFFERING IN THEIR RELATION TO THE PAYLOAD GRID. 64-BIT VARIANT, COCO VAL2017, N = 1000.

| Attack | Grid | Severity (dB) | BER |
|---|---|---|---|
| brightness 0.5 | preserved | 12.24 | 0.0001 |
| crop to 50% | displaced | 12.22 | 0.4931 |
| solarize 0.5 | preserved | 11.39 | 0.0225 |
| rotate 30 deg | displaced | 11.70 | 0.4952 |
| cutout 20% | preserved | 18.66 | 0.0896 |
| rescale 0.7 | displaced | 29.44 | 0.4748 |

Measured severity is recorded alongside the training status, and for the same reason. A bit error rate quoted against a nominal setting conveys less than it appears to: the same setting produces different amounts of degradation depending on the implementation and on image content, as the blur kernel above illustrates, and encoders with rate control allocate bits by content, so a fixed quality parameter is not a fixed distortion. Every row of every table therefore reports the measured severity of the attack alongside the error rate, as the peak signal-to-noise ratio of the attacked image against the watermarked one. Robustness benchmarks for watermarking have adopted the same convention of reporting detection performance jointly with the induced degradation [10].

### D. REPRODUCIBILITY AND TEST SETS

Stochastic attacks are seeded per image and per attack, so a run is reproducible to the bit and two models compared on the same suite are compared on identical pairs, which permits paired tests between them. Cover images and payloads are paired by file name and payload bits are generated deterministically from the image path, so a given image always carries the same message on every machine.

Two test sets are used in a single pass. BSDS300, of one hundred images, provides continuity with earlier work [32]; a subset of one thousand images from COCO val2017 provides comparability with the reference systems discussed in Section III, which report on that corpus. Both are prepared as native center crops at the working resolution, so no resampling occurs inside the evaluation and the two sets differ in content rather than in preparation.

The two sets agree on error rates. Across the whole suite and all four capacities, the mean absolute difference between them is 0.0046, with a maximum of 0.032 on any single attack; the difference in transparency reported in Section IV-A does not carry over into the readings. What follows in Section V is therefore a property of the scheme rather than of the corpus.

## V. RESULTS: THREE LIMITING MECHANISMS

This section reports three mechanisms that limit robustness in the class of schemes described in Section I. Each is stated first as a property of the class of designs and then measured on the instance of Section IV, so that the argument does not depend on the number of systems examined.

### A. DESYNCHRONIZATION OF THE GRID

Where the payload is indexed by position, extraction reads element k from the region where element k was written. A transformation that moves that region breaks the correspondence between writing and reading, and it does so whether or not the embedded signal survives the transformation intact. The loss is then a loss of synchronization, not of signal strength, and the two behave differently: attenuation degrades extraction gradually, while displacement removes the basis on which extraction operates at all. This distinction belongs to the class, and it yields three predictions that can be checked independently of one another.

The first prediction is that the measured severity of an attack should not predict its effect on the payload. If displacement is what matters, then a mild transformation that moves the grid should be more damaging than a violent one that leaves it in place. Table 5 pairs attacks matched on measured severity and differing only in their relation to the grid. Brightness at 0.5 and a crop to half the area sit 0.02 dB apart in severity and differ by a factor of more than four thousand. Solarization and rotation by 30 degrees sit 0.31 dB apart and differ by a factor of twenty-two. The third pair inverts the severity ordering: cutout is 10.8 dB more severe than rescaling to 0.7, and damages the payload five times less. Across the suite, severity and damage are unrelated: the rank correlation between them is −0.06 over thirty-one attacks. Within each pair it is the relation to the grid that separates the members, and it does so completely.

The second prediction is quantitative. If the payload occupies disjoint regions, removing a fifth of the image should corrupt a fifth of it and nothing more. Blanked blocks decode at chance, so the expected error rate is half of one fifth, or 0.10, at every capacity. The measured values are 0.0896, 0.0956, 0.0973 and 0.1016 across the four payloads, converging on that limit as the block grid becomes finer. Nothing spills into the blocks that were not removed.

The third prediction concerns scaling. Tolerance to displacement should be bounded by block size, because a shift small relative to a block leaves most of it in place while the same shift on a block a quarter the size does not. Table 6 gives rotation by 10 degrees and a crop retaining four fifths of the area, the only grid-displacing attacks mild enough to leave the measurement unsaturated. Rotation moves the error rate from 0.221 to 0.500 as the block shrinks from 32 to 2 pixels, and the crop from 0.183 to 0.499 over the same range. A rotation the extractor tolerates at 32 pixels is indistinguishable from guessing at 2.

TABLE 6
BIT ERROR RATE UNDER GRID-DISPLACING ATTACKS AS A FUNCTION OF BLOCK SIZE. COCO VAL2017, N = 1000.

| | **64 bits** | **256 bits** | **1024 bits** | **16384 bits** | **4096 bits** |
|---|---|---|---|---|---|
| Block size (px) | 32 | 16 | 8 | 2 | 8 |
| Bits per block | 1 | 1 | 1 | 1 | 4 |
| rotate 10 deg | 0.2210 | 0.4395 | 0.4907 | 0.4999 | 0.4979 |
| crop to 80% | 0.1832 | 0.3728 | 0.4609 | 0.4991 | 0.4883 |
| no attack | 0.0001 | 0.0001 | 0.0001 | 0.0006 | 0.0278 |

TABLE 7
BIT ERROR RATE FOR THREE EXTRACTORS FINE-TUNED ON DIFFERENT DISTORTIONS, ONE OF THEM GEOMETRIC. 64-BIT VARIANT, COCO VAL2017, N = 1000.

| **Attack** | **Grid** | **Base** | **Photometric** | **Geometric** |
|---|---|---|---|---|
| rotate 5 deg (trained) | displaced | 0.0212 | 0.0202 | 0.0184 |
| rotate 10 deg | displaced | 0.2210 | 0.2208 | 0.2179 |
| rotate 30 deg | displaced | 0.4952 | 0.4956 | 0.4920 |
| rescale 0.9 (trained) | preserved | 0.2344 | 0.3690 | 0.2052 |
| crop to 80% | displaced | 0.1832 | 0.1985 | 0.1718 |
| contrast 2.0 (control) | | 0.0777 | 0.0827 | 0.0940 |
| Gaussian blur k7 (control) | | 0.4889 | 0.0746 | 0.2371 |

In those four variants block size and capacity move together, so the scaling just described admits a competing reading, namely that a larger payload is simply harder to recover. The last column of Table 6 separates the two. That variant carries four bits in the grey level of each block rather than one in its sign, so 4096 bits sit on the same 8-pixel grid as the 1024-bit variant. Quadrupling the payload at constant block size moves rotation from 0.4907 to 0.4979 and the crop from 0.4609 to 0.4883, while reducing the block from 32 to 8 pixels moves rotation from 0.221 to 0.491.

Two limits of that comparison should be stated. The variant carrying four bits per block is weaker at its baseline, reaching 0.0278 on clean images against 0.0001 to 0.0006 for the others. That is the cost of resolving four grey levels instead of a sign, not a deficiency of training, and it shows on the grid-preserving attack, where rescaling by 1.5 leaves 0.0007 at 1024 bits and 0.3505 at 4096. The second limit is saturation. At a block of 8 pixels both members of the pair already read at chance, so their agreement carries less weight than a separation would. The series measures how far tolerance falls, while the control, limited as it is, is what rules out the payload as the cause. Block geometry, not the number of bits, is the governing quantity.

The obvious remedy is to train the extractor on displaced images. Three extractors were compared at 64 bits, reading images from the same frozen encoder and differing only in the distortions used to train them. The first is the base extractor, which sees no distortion during training. The second was fine-tuned on photometric distortions alone, color jitter, Gaussian blur and solarization. The third was fine-tuned on a profile adding rotation up to 10 degrees, scaling between 0.8 and 1.0 and perspective warping. For this experiment the suite of Section IV-B was extended by two attack points, a rotation of 5 degrees and a rescaling to 0.9. Both lie inside the geometric training range, and they were added so that the mechanism is examined where training has the best chance of overcoming it, and not only at severities where no method would be expected to succeed. The last two rows are photometric controls, and their role is to show that the fine-tuning procedure itself is effective. Table 7 reports the result.

Even at the point most favorable to it, a rotation of 5 degrees inside its training range, the geometric extractor improves on the base extractor by only 0.003, from 0.0212 to 0.0184, and the margin at 10 degrees is the same. Beyond it, at 30 degrees, all three read at chance. Training on displacement does not confer tolerance to displacement, not even within its own training range.

The added points also reveal a cost that the standard suite hides. Photometric fine-tuning, which the control rows show to be effective, degrades both mild geometric attacks, by 0.135 on rescaling to 0.9 and by 0.015 on a crop retaining four fifths. Both regressions replicate on BSDS300, and the geometric extractor reverses both. Rescaling preserves the grid while cropping displaces it, so the larger cost falls where synchronization is not at stake: what photometric fine-tuning costs is the margin with which each block is read once resampling has softened it.

The gain is offset by losses in other classes. Contrast doubling rises from 0.0777 at the base to 0.0940, and the transfer to blur that photometric training produces is only partly retained, reaching 0.2371 against 0.0746.

The mechanism is therefore not addressed by training the extractor harder. It is addressed by restoring the correspondence between writing and reading, which is a synchronization problem and has a literature of its own. The same limit appears in schemes built on entirely different foundations. Watermarks that bind payload bits to spatially indexed carriers in the latent space of a diffusion model are reported to fail under rotation-induced misalignment [13]. In that family the strongest geometric invariance belongs to zero-bit constructions, which carry no message to index [12].

TABLE 8
BIT ERROR RATE FOR FOUR EXTRACTORS FINE-TUNED ON DIFFERENT DISTORTIONS, TWO OF THEM COMPRESSION. COCO VAL2017, N = 1000.

| Attack | Base | Photometric | Codec | JPEG |
|---|---|---|---|---|
| no attack | 0.0001 | 0.0001 | 0.0001 | 0.0001 |
| H.264 CRF 16 (trained) | 0.1091 | 0.1318 | 0.2033 | 0.1885 |
| H.264 CRF 20 (trained) | 0.2114 | 0.2329 | 0.3266 | 0.3042 |
| H.264 CRF 23 (trained) | 0.2987 | 0.3102 | 0.4001 | 0.3795 |
| H.264 CRF 30 | 0.4453 | 0.4457 | 0.4830 | 0.4773 |
| JPEG 75 (trained) | 0.2879 | 0.3291 | 0.3723 | 0.3953 |
| JPEG 50 | 0.3924 | 0.4293 | 0.4631 | 0.4720 |
| Gaussian blur k3 | 0.0164 | 0.0001 | 0.0002 | 0.0002 |
| contrast 2.0 | 0.0777 | 0.0827 | 0.0816 | 0.0927 |
| salt-and-pepper 5% | 0.4047 | 0.0751 | 0.4798 | 0.4465 |

ChunkySeal, the highest-capacity system in the literature, reports robustness over rotations of at most 10 degrees and crops retaining between 77 and 95 percent of the area [11]. That band stops well short of the range measured here, and is consistent with the limit described above.

### B. COMPRESSION AS A TRAINING TARGET

Training extraction separately from embedding removes a technical requirement, as Section III-C describes. The distortion is an input to a stage rather than a layer on a gradient path, so a real codec can be placed in training without a differentiable substitute. MBRS had to construct one precisely because in their single-stage design the distortion lies on the encoder's gradient path. This section measures what that freedom is worth.

Four extractors were compared, all reading images from the same frozen watermark encoder, so the images they see are identical. The first is the base extractor of Section V-A, which sees no distortion. The other three were fine-tuned from it under the same settings, seed and schedule, and differ only in the distortions used for fine-tuning. All three share a photometric core: color jitter with probability 0.8, covering brightness between 0.6 and 1.4, contrast and saturation between 0.5 and 2.0 and hue within plus or minus 0.25; Gaussian blur with probability 0.4 at kernel 3 and standard deviation between 1.0 and 2.5; and solarization with probability 0.2 at threshold 0.5. The second uses that core alone and is the photometric extractor of Section V-A. The third adds a real H.264 encoder with probability 0.5 at constant rate factors 16 to 23, and the fourth real JPEG at qualities 75 to 95 in its place. Table 8 reports the result.

The experiment is run at the lowest capacity examined, and that choice is deliberate. Blocks are largest and reading margins widest there, and the compression class is correspondingly least damaging, rising monotonically with capacity from 0.3503 at 64 bits to 0.4913 at 16384—the Appendix gives the full series. This is therefore the condition most favorable to the hypothesis under test, and a failure here implies a failure in every harder regime.

Two details matter for reading Table 8. The first concerns blur. The blur seen during training uses a kernel of 3, and the blur row of that table reports the same operating point, not the kernel of 7 used in the suite of Section IV-B and in the blur control of the geometric comparison above. The row is a control on what the photometric extractor was trained for, which is what makes its improvement interpretable. The second concerns the codec. The suite was extended here in the same way and for the same reason as in Section V-A, by two further points, constant rate factors 16 and 20, which lie inside the training range of the codec-trained extractor.

The first reading of Table 8 is the one that settles the question. Both extractors trained on compression are worse than the base extractor at exactly the operating points they were trained on. The codec extractor, trained at constant rate factors 16 to 23, loses at all three of them to an extractor that has never seen a codec, by 86 percent at rate factor 16, where the error rate rises from 0.1091 to 0.2033. The JPEG extractor, trained at quality 75 and above, reaches 0.3953 at quality 75 against 0.2879 for the base extractor, 37 percent worse. Neither result is an artifact of training budget. Both regressions are significant under a paired test over the thousand images. The codec extractor loses 0.094 at rate factor 16, with a 95 percent interval of 0.089 to 0.100, and the JPEG extractor loses 0.107 at quality 75, with an interval of 0.103 to 0.112. Both compression profiles were stopped by the patience criterion, which fires only when a run stops improving. Training on compression degrades the reading of the very points used in training.

The second reading rules out the obvious objection, that the extractor simply cannot learn anything at this capacity. On Gaussian blur the photometric extractor improves on the base extractor by two orders of magnitude, from 0.0164 to 0.0001.

More telling still, on salt-and-pepper noise, which appears in no training profile at all, the same extractor improves from 0.4047 to 0.0751, a factor of five obtained purely by generalization from other photometric distortions. The network learns and generalizes, but it does not learn compression.

The third reading is the one that turns a null result into a cost. On that same salt-and-pepper attack the codec and JPEG extractors reach 0.4798 and 0.4465, which is worse than the base extractor that was never fine-tuned at all. Training on compression does not merely fail to confer robustness to compression. It destroys the robustness that photometric fine-tuning would otherwise have produced elsewhere.

TABLE 9
BIT ERROR RATE ON CLEAN IMAGES AS A FUNCTION OF THE EMBEDDING-STRENGTH FACTOR. COCO VAL2017, N = 1000.

| Payload | 0.5 | 0.75 | 1.0 | 1.25 | 1.5 | 2.0 |
|---|---|---|---|---|---|---|
| 64 | 0.0434 | 0.0022 | 0.0001 | 0.0002 | 0.0009 | 0.0150 |
| 256 | 0.1323 | 0.0104 | 0.0001 | 0.0003 | 0.0044 | 0.0711 |
| 1024 | 0.1241 | 0.0166 | 0.0001 | 0.0008 | 0.0100 | 0.0505 |
| 16384 | 0.3165 | 0.1277 | 0.0006 | 0.0272 | 0.0799 | 0.1446 |

TABLE 10
THE USABLE EMBEDDING-STRENGTH WINDOW AT AN ERROR RATE OF 0.001 ON CLEAN IMAGES, AND THE PSNR IT SPANS. COCO VAL2017, N = 1000.

| Payload (bits) | Block (px) | Usable settings | Count | Transparency (dB) |
|---|---|---|---|---|
| 64 | 32 | 1.0 - 1.5 | 3 | 35.6 - 39.1 |
| 256 | 16 | 1.0 - 1.25 | 2 | 37.6 - 39.5 |
| 1024 | 8 | 1.0 - 1.25 | 2 | 36.5 - 38.4 |
| 16384 | 2 | 1.0 only | 1 | 36.1 |

The same contrast appears inside a single run. In the codec run the validation blur metric falls by three orders of magnitude over 62 epochs, while the codec metric stays flat. It starts below chance level, so the margin to improve existed and went unused. The comparison is therefore not an artifact of how the four extractors were selected.

The following is interpretation rather than measurement. A loss term whose target is not learnable still contributes gradient, and that gradient is noise with respect to the objective. An optimizer receiving it drifts toward solutions better on the learnable classes and worse on the unlearnable one, which is what the third reading records.

MBRS [9] anticipated this ceiling, observing that an encoder which cannot obtain information about decoding under JPEG distortion will not perform well enough. The measurements here carry that observation to the extraction side, where pushing against the same ceiling proves counterproductive rather than merely insufficient. They also confirm the objection that decoupling yields only a locally optimal solution, and put a number on what freezing the embedder costs. The scope of that claim should be read precisely. What was measured is one extraction architecture in three training variants, not extraction architectures in general. The generalization runs through the mechanism rather than a survey of architectures. The embedder is frozen and never observes the distortion, so the information is already absent from what reaches extraction, and no reading head recovers what is not present at its input. That is a prediction about other architectures, not a measurement of them.

One design implication follows directly. Robustness to a codec must be acquired somewhere, and the measurement removes extraction as the place. That leaves embedding, and with it the differentiable codec model that decoupling was adopted to escape. It costs transparency too, since the allocation that survives compression is not the one that minimizes visibility, as Section II-A describes. This is a conclusion from measurement, and it reverses the direction in which the decoupled schemes of Section III-C were expected to be extended.

### *C. THE USABLE EMBEDDING-STRENGTH WINDOW*

Embedding strength is the oldest lever in watermarking, and in the designs studied here it is the only one that remains available after training. The watermarked image is composed as the cover plus the luminance difference produced by the encoder, so that difference can be scaled at embedding time, with no retraining and no change to the extractor. The factor cannot be moved freely in either direction. Below some value the difference is too weak to be read. Above some value it becomes visible, and the composed image begins to exceed the valid range of pixel values, where the excess is cut off and the difference stops growing. The interval between the two bounds is the usable window.

The reason belongs to the class rather than to the network. Detectability of a payload element depends on the total signal carried by its block, which is amplitude multiplied by area. As blocks are subdivided, area falls, so the amplitude required for the same detectability rises, and the composed image sits closer to the top of the valid range before any scaling is applied at all. Headroom above the trained operating point therefore shrinks with block size, and the margin below it shrinks for the same reason, because there is less signal to give away. The window should close from both directions at once.

Table 9 gives the bit error rate on clean images as the factor is varied from 0.5 to 2.0. The response is not monotone in the factor. At every capacity the error rate rises on both sides of the trained value, which is what separates this mechanism from a simple threshold. The payload is lost by being embedded too strongly just as it is lost by being embedded too weakly.

The narrowing itself is visible without choosing any threshold. At each of the three settings adjacent to the trained value, 0.75, 1.25 and 1.5, the error rate rises monotonically with capacity. The same displacement from the trained value costs between one and two orders of magnitude more at 16384 bits than at 64. At the extreme settings of 0.5 and 2.0 the ordering does not hold, but the payload is unreadable at every capacity there, so the comparison is between failures rather than between working operating points.

A threshold turns the ordering into a count. Taking an error rate of 0.001 on clean images as the condition for readability, the number of usable settings falls from three at 64 bits to two at 256 and 1024 and one at 16384. Table 10 gives those windows with the transparency each spans. At the highest capacity the trained value is the only setting that works, since both its neighbors lie far above the threshold. The count depends on that threshold and on the six-point grid, which is why the ordering above is the more robust statement. On BSDS300, whose images are easier, the same threshold admits one additional setting at 64 and at 256 bits.

What is lost with the window is the classical trade-off itself, between payload, imperceptibility and robustness. Raising the factor is the standard way of increasing robustness at a cost in transparency, and where the window is wide it works. At 64 bits moving from 1.0 to 1.5 costs 3.5 dB and reduces the error rate under JPEG at quality 90 from 0.1113 to 0.0803, under JPEG at quality 75 from 0.2879 to 0.2460, and under H.264 at constant rate factor 23 from 0.2987 to 0.2623. Where the window has closed to a single setting the lever does not exist in either direction, and a designer working at 16384 bits has one operating point rather than a curve.

The control variant of Section V-A places this in a wider context. Carrying four bits in the grey level of an 8-pixel block rather than one in its sign, it reads at 0.0278 on clean images even at the trained value, so under the same criterion it has no usable window at all. The reading margin that the window measures is consumed either by smaller blocks or by more bits per block, and the two draw on it in the same way.

Two qualifications limit how far this should be read. The threshold of 0.001 is a convention, and a scheme tolerating a higher error rate would measure a wider window at every capacity, which is why the ordering of the curves rather than the count carries the argument. And unlike the two mechanisms preceding it, this one is not a failure under any particular attack. It is the loss of a degree of freedom that the designer of a low-capacity scheme takes for granted.

## VI. DISCUSSION AND DESIGN IMPLICATIONS

The question posed in Section I can now be answered. Raising capacity narrows the margin on which extraction relies, but it is not what limits robustness. In each of the three cases the governing quantity is block geometry or the training design rather than the size of the payload. Quadrupling the payload at constant block size leaves the error rate under rotation almost unchanged, while quartering the block moves it from 0.221 to 0.491. The compression limit is fully present at the lowest capacity measured, where JPEG at quality 75 already gives 0.2879. What separates a distortion the scheme survives from one it does not is the class the distortion belongs to, and capacity is the axis along which the margins narrow rather than the reason they run out.

The three mechanisms are also disjoint. They damage different things, they scale differently with capacity, and each requires a different response. Geometric transformation displaces the reference against which the payload is read, leaving the signal intact and the reading impossible. Compression removes the signal itself, and does so in a way the extraction stage cannot be taught to compensate. The embedding-strength window is not a failure under any attack at all, but the loss of a degree of freedom that the designer of a low-capacity scheme assumes is available. Treating these as three manifestations of a single quantity called robustness is what makes them look like a trade-off against capacity.

All three arise in the same place. They are already present in the representation that reaches the extraction stage: the grid has been displaced, the signal has been quantized away, or the amplitude was never available to begin with. This has a practical consequence, because the default response to weak robustness in this field is to add distortions to training and continue. Sections V-A and V-B measured what that produces when the distortion is not learnable, once for geometry and once for compression, and the two agree. Adding the unlearnable class gains almost nothing at the operating points it was added for, and it degrades the classes that were learnable, so the cost falls on robustness the extractor would otherwise have acquired. Additional training effort on the extraction side is not a weak remedy here. It is the wrong axis. Section V-B states the scope of that claim, and states it as a prediction, falsifiable by any architecture that recovers a payload the frozen embedder never protected.

### A. GENERALITY OF THE THREE MECHANISMS

Each mechanism has a different reach, and it is worth stating them separately rather than as one claim. Grid desynchronization applies to any scheme that indexes the payload spatially, whatever the network. The codec limit applies to any scheme that trains extraction separately from a frozen embedder, whatever the payload arrangement. The narrowing embedding-strength window applies to any scheme whose embedding strength can be scaled after training, which is the case whenever the watermarked image is composed as the cover plus a residual.

The first of the three has independent confirmation outside the family measured here. As Section V-A notes, watermarks that index payload bits by position in the latent space of a diffusion model fail under rotation-induced misalignment. The carrier is derived in an entirely different way and the failure is the same, because what produces it is the indexing rather than the derivation. The strongest geometric invariance reported anywhere belongs instead to zero-bit constructions, where invariance follows from the geometry of a fixed pattern rather than from training [12], [14], and that same fixed geometry is what bounds the payload. Capacity is obtained by localizing information, and localization is precisely what a geometric transformation disturbs.

This refines the classical trade-off between payload, imperceptibility and robustness rather than contradicting it. Section V-C measures it directly, in the closing of the embedding-strength window. It governs only the amplitude budget, since a displaced grid is a failure of correspondence rather than of signal energy, and the codec limit belongs to the training decomposition and is present at the lowest capacity measured. What remains is a narrower trade-off, between capacity and geometric invariance, since capacity comes from localizing the payload and invariance from not depending on where it sits. No scheme in the literature has both.

### B. IMPLICATIONS FOR DESIGN

Each mechanism points somewhere different, and none of the three points at the extraction stage. For grid desynchronization Section V-A already names the remedy, a synchronization mechanism rather than wider training coverage. What the measurements add is its specification, since the tolerance available at each block size is what such a mechanism would have to supply.

Compression points at the embedder. Section V-B derives that implication from the measurement, and what belongs here is its consequence for the strategies of Section III-C. Decoupling was adopted because it removes the technical requirement for a differentiable model of the distortion. The measurement says that model is needed anyway, on the other side of the frozen boundary, so the requirement returns rather than disappearing.

The embedding-strength window points at the operating point itself. Where the window is wide, the strength can be chosen after training to suit a deployment. Where it has closed, there is a single operating point, and the transparency-robustness curve that the field reports as a design space does not exist. A designer working at the upper end of the capacity range therefore has one fewer parameter than the literature assumes, and the remaining axis is the carrier rather than its amplitude. Chrominance, left untouched for the reasons given in Section IV-A, is unused capacity in exactly that sense.

### C. LIMITATIONS

The measurements were made at a single resolution, 256 by 256 pixels, and block sizes are stated in pixels rather than as a fraction of the image. The mechanism of Section V-A depends on the ratio of displacement to block size, so the numbers transfer to other resolutions only after that ratio is preserved. One family of embedding architectures was used throughout, and while Section V-B is stated as a property of decoupled training rather than of a network, it was measured on one. Codecs were run in the default configuration of a single implementation. Rate control and preset selection affect the severity of the distortion, and the absolute figures would move under a different configuration, although the comparison between the four extractors of Table 8 would not, since all four were measured under identical settings. Attacks that regenerate content rather than degrade it were excluded for the reason given in Section II-D, and the question of whether a payload of this kind can survive a semantics-preserving resynthesis remains open.

## VII. CONCLUSION

This paper asked what limits robustness in deep image watermarking, and whether the limits move with payload capacity. The answer separates into two parts Robustness is not one property but at least three disjoint mechanisms, and capacity governs only one of them. It is the axis along which the margins narrow and along which the mechanisms become measurable, which is a different statement and leads to different engineering.

The three were identified and measured across payloads spanning well beyond the range for which the strategies reviewed in Section III were designed. Displacement of the payload grid under geometric transformation is bounded by block size and is a failure of synchronization rather than of signal strength. Its severity is unrelated to how violent the attack looks. Codec-induced distortion cannot be learned by an extraction stage trained against a frozen embedder, and attempting it degrades the reading of the very operating points used in training while destroying robustness acquired elsewhere. The usable embedding-strength window narrows with capacity and closes entirely at the upper end, removing the transparency-robustness trade-off that lower-capacity schemes take for granted.

None of the three is removed by training the extraction stage harder, because all three arise before it, in the representation extraction receives at its input. None of the three is removed by training the extraction stage harder, because all three arise before it, in the representation extraction receives at its input. The boundary runs along the class of distortion rather than along the capacity axis, and the classes that fail do so at every capacity, including the lowest.

The conclusions are properties of a class of designs rather than of one system, and each has a different reach: the first covers any scheme that indexes its payload spatially, the second any scheme that separates the training of extraction from embedding, the third any scheme whose embedding strength can be scaled after training. The evaluation protocol introduced in Section IV records for every attack an explicit parameter, a measured severity, a three-valued training-status annotation and, for most attacks, the published system that reports the same operating point. That is what makes claims of this kind checkable rather than declarative, and what allows results obtained under it to be set beside published figures. It is offered for use beyond the present work.

Two of the three mechanisms point at parts of the system that this paper deliberately held fixed: synchronization ahead of extraction, and the embedder itself. In a practical deployment at these capacities the decisive quantity is not the error rate but the amount of payload that can be relied on, since a license or a signature of the kind described in Section I verifies only as a whole. How much of a payload survives once these limits are accepted, and what redundancy is needed to make it usable, is a separate question and the subject of continuing work.

## APPENDIX

TABLE 11
BIT ERROR RATE OF THE FIVE BASE VARIANTS UNDER THE FULL ATTACK SUITE OF SECTION IV-B, AT THE TRAINED EMBEDDING STRENGTH. COCO VAL2017, N = 1000.

| Attack | Grid | Severity (dB) | 64_1 | 256_1 | 1024_1 | 16384_1 | 4096_4 |
|---|---|---|---|---|---|---|---|
| **Control** | | | | | | | |
| no attack | preserved | n/a | 0.0001 | 0.0001 | 0.0001 | 0.0006 | 0.0278 |
| **Point operations** | | | | | | | |
| brightness 0.5 | preserved | 12.23 | 0.0001 | 0.0003 | 0.0003 | 0.0054 | 0.2248 |
| contrast 0.5 | preserved | 18.60 | 0.0001 | 0.0003 | 0.0004 | 0.0070 | 0.2516 |
| hue 0.25 | preserved | 22.87 | 0.0001 | 0.0004 | 0.0010 | 0.0188 | 0.1125 |
| saturation 2.0 | preserved | 26.90 | 0.0017 | 0.0019 | 0.0029 | 0.0103 | 0.0767 |
| histogram equalize | preserved | 17.15 | 0.0038 | 0.0183 | 0.0311 | 0.1143 | 0.3563 |
| solarize 0.5 | preserved | 11.33 | 0.0225 | 0.0563 | 0.0823 | 0.1538 | 0.3266 |
| brightness 1.5 | preserved | 14.67 | 0.0507 | 0.0667 | 0.0855 | 0.1072 | 0.2749 |
| contrast 2.0 | preserved | 17.14 | 0.0777 | 0.1033 | 0.1240 | 0.1744 | 0.3643 |
| **Filtering and noise** | | | | | | | |
| salt-and-pepper 1% | preserved | 24.87 | 0.0044 | 0.0440 | 0.1114 | 0.1619 | 0.4426 |
| Gaussian noise 0.03 | preserved | 30.58 | 0.1939 | 0.2992 | 0.3394 | 0.4136 | 0.4612 |
| Gaussian noise 0.06 | preserved | 24.69 | 0.3385 | 0.4151 | 0.4283 | 0.4659 | 0.4859 |
| salt-and-pepper 5% | preserved | 17.87 | 0.4047 | 0.4515 | 0.4228 | 0.4444 | 0.4910 |
| Gaussian blur k7, sigma 2.0 | preserved | 24.42 | 0.4889 | 0.4962 | 0.4940 | 0.5001 | 0.4998 |
| Gaussian blur k7, sigma 1.0 | preserved | 27.46 | 0.4955 | 0.4914 | 0.4890 | 0.4948 | 0.5001 |
| **Compression** | | | | | | | |
| JPEG 90 | preserved | 37.48 | 0.1113 | 0.1934 | 0.3668 | 0.4527 | 0.4494 |
| JPEG 75 | preserved | 33.27 | 0.2879 | 0.4008 | 0.4793 | 0.4939 | 0.4907 |
| H.264 CRF 23 | preserved | 31.53 | 0.2987 | 0.4233 | 0.4777 | 0.4923 | 0.4919 |
| H.265 CRF 28 | preserved | 30.54 | 0.3356 | 0.4336 | 0.4842 | 0.4948 | 0.4952 |
| JPEG 50 | preserved | 30.72 | 0.3924 | 0.4707 | 0.4967 | 0.4983 | 0.4988 |
| JPEG 30 | preserved | 29.02 | 0.4431 | 0.4912 | 0.4985 | 0.4993 | 0.4999 |
| H.264 CRF 30 | preserved | 27.47 | 0.4453 | 0.4912 | 0.4988 | 0.4994 | 0.4996 |
| JPEG 10 | preserved | 25.43 | 0.4882 | 0.4974 | 0.4997 | 0.4999 | 0.4999 |
| **Spatial** | | | | | | | |
| cutout 20% | preserved | 18.63 | 0.0896 | 0.0956 | 0.0973 | 0.1016 | 0.3568 |
| crop to 80% | displaced | 14.54 | 0.1832 | 0.3728 | 0.4609 | 0.4991 | 0.4883 |
| crop to 50% | displaced | 12.20 | 0.4931 | 0.4992 | 0.5024 | 0.5000 | 0.4997 |
| **Geometric** | | | | | | | |
| rescale 1.5 | preserved | 34.83 | 0.0001 | 0.0002 | 0.0007 | 0.0235 | 0.3505 |
| rotate 10 deg | displaced | 13.78 | 0.2210 | 0.4395 | 0.4907 | 0.4999 | 0.4979 |
| rescale 0.7 | displaced | 29.31 | 0.4748 | 0.4577 | 0.4862 | 0.4867 | 0.4972 |
| rotate 30 deg | displaced | 11.68 | 0.4952 | 0.5002 | 0.4995 | 0.5000 | 0.4997 |
| **Composite** | | | | | | | |
| H.264 30, brightness 0.5, JPEG 75 | preserved | 12.05 | 0.4500 | 0.4881 | 0.4981 | 0.4997 | 0.5000 |
| H.264 30, brightness 0.5, crop 50% | displaced | 10.16 | 0.4948 | 0.4997 | 0.5002 | 0.5000 | 0.5001 |

Table 11 reports the bit error rate of the five base variants under the full attack suite of Section IV-B, on COCO val2017 with one thousand images per point and the embedding strength at its trained value. The severity column gives the mean peak signal-to-noise ratio between the watermarked image and its attacked version, averaged over the five variants, which differ by at most 0.68 dB on any attack. The grid column records whether the attack moves the payload block grid, in the sense used in Section V-A.

Attacks are grouped by the classes of Table 4 and ordered within each class by their effect at 64 bits.

The compression class cited in Section V-B is the mean of the eight compression rows, which runs 0.3503, 0.4252, 0.4752 and 0.4913 across the four single-bit variants. Results on BSDS300 are not reproduced here, since Section IV-D reports the mean absolute difference between the two test sets over this same suite.

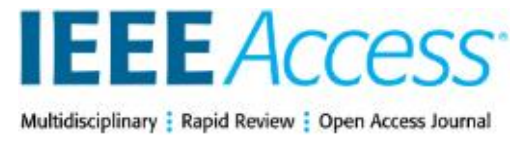

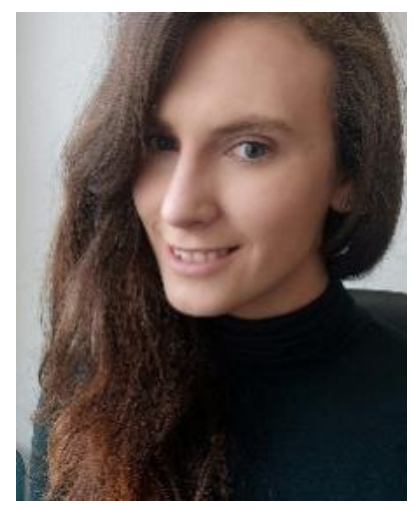

**MARTA BISTROŃ** received the M.Sc. degree in energy engineering from the Military University of Technology (MUT), Poland, in 2018. Since 2019, she has been pursuing the Ph.D. degree in technical informatics and telecommunications at MUT. Since 2021, she has been working as a Research and Teaching Assistant with the Institute of Communication Systems, Faculty of Electronics, MUT. Her research interests focus on computer vision, particularly multimedia signal processing, image and video watermarking, and super-resolution methods using deep learning techniques.

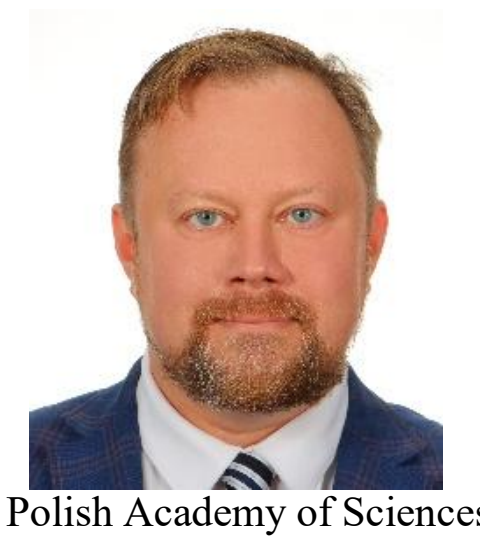

**ZBIGNIEW PIOTROWSKI** received the Ph.D. and D.Sc. degrees in telecommunications from the Military University of Technology (MUT), Poland, in 2005 and 2013, respectively. He is currently a Professor with MUT and the Leader of the Research and Development Group with the Institute of Communication Systems, Faculty of Electronics, MUT. He also serves as an expert on the Committee on Electronics and Telecommunications of the Polish Academy of Sciences. His research interests focus on the application of artificial intelligence and digital signal processing (DSP) in autonomous systems, radio communication systems, and multimedia.